\documentclass[11pt]{article}

\usepackage[margin=1in]{geometry}
\usepackage{amsmath,amssymb,amsthm,mathtools}
\usepackage{booktabs}
\usepackage{enumitem}
\usepackage[hidelinks]{hyperref}

\newtheorem{theorem}{Theorem}

\newcommand{\Or}{\mathcal{O}}
\newcommand{\ip}[2]{\left\langle #1,#2\right\rangle_F}
\newcommand{\tr}{\operatorname{tr}}

\title{Local Maxima of the Entrywise $\ell_4$ Norm\\
on the Orthogonal Group}
\author{Dian Jin\\
\small Department of Statistics and Data Science\\
\small National University of Singapore\\
\small \texttt{dianjin@nus.edu.sg}}
\date{}

\begin{document}
\maketitle

\begin{abstract}
We classify the local maximizers of the entrywise fourth-power objective
\[
Q\longmapsto \lVert Q\rVert_4^4
=\sum_{i,j=1}^r q_{ij}^4
\]
over the real orthogonal group $\mathcal O(r)$.  We prove that the signed
permutation matrices are the only local maximizers, and hence the only global
maximizers, in every dimension.  More strongly, every other stationary point
has an explicit rank-two tangent direction with strictly positive second
variation.  The proof is based on a maximum-entry pivot for the orthostochastic
matrix $Q^{\circ2}$: the associated full Riemannian Hessian can be evaluated
exactly and is positive at a largest nonunit squared entry.  The argument is
self-contained and handles zeros, repeated magnitudes, reducible support, and
Hadamard-type stationary points.
\end{abstract}


Throughout, powers marked by ``$\circ$'' are entrywise.  In particular,
$Q^{\circ 2}=(q_{ij}^2)$ and $Q^{\circ 3}=(q_{ij}^3)$.  The norm in the
objective is also entrywise:
\[
 f_r(Q)=\lVert Q\rVert_4^4=\sum_{i,j=1}^r q_{ij}^4,
 \qquad Q\in\Or(r).
\]

\section{Problem formulation and main result}

The orthogonal group is compact but nonconvex, and entrywise norm objectives
have a markedly different geometry from spectral objectives.  The squared
Frobenius norm is constant on $\mathcal O(r)$, whereas the fourth-power
objective
\[
 f_r(Q)=\lVert Q\rVert_4^4,
 \qquad Q\in\mathcal O(r),
 \tag{1.1}
\]
rewards concentration of mass in individual entries.  This paper determines
its complete local-maximality landscape over the real square orthogonal group.

The signed permutation matrices are exactly the local maximizers, and the
maximum value is $r$.  More strongly, every stationary point that is not a
signed permutation has an explicit feasible rank-two direction with strictly
positive second variation.  Thus it has a strict ascent direction for the
maximization problem.

\begin{center}
\begin{tabular}{@{}llll@{}}
\toprule
Dimension & Local maximizers & Maximum & Other stationary points \\
\midrule
$r=1$ & $[1]$ and $[-1]$ & $1$ & none \\
$r=2$ & signed permutations only & $2$ & strict ascent direction exists \\
$r\ge3$ & signed permutations only & $r$ & strict ascent direction exists \\
\bottomrule
\end{tabular}
\end{center}

Here ``strict ascent direction'' means that the largest eigenvalue of the
constrained Hessian is positive.  This is sometimes called the strict-saddle
property with the maximization sign convention.  It does not assert that the
Hessian is necessarily indefinite: a local minimum also has positive
curvature directions.

\section{Related work}

The Riemannian geometry of optimization with orthogonality constraints is
classical; standard treatments include the geometric analysis of the Stiefel
and Grassmann manifolds by Edelman, Arias, and Smith~\cite{EdelmanAriasSmith1998}
and the monograph of Absil, Mahony, and Sepulchre~\cite{AbsilMahonySepulchre2008}.
We use this framework only to derive the exact stationarity condition and the
second variation along $Qe^{t\Omega}$.

Our problem also belongs to the study of entrywise $\ell_p$ norm landscapes on
the orthogonal group.  In particular, the almost-Hadamard program investigates
local extrema of such norms, especially in connection with the $\ell_1$ norm,
Hadamard matrices, and more general exponents; see Banica and
Nechita~\cite{BanicaNechita2013}.  The $p=4$ maximization problem has the
opposite global extremizers from the normalized Hadamard configurations:
signed permutations maximize $\lVert Q\rVert_4^4$.  This deterministic
maximization problem was studied by Zhai et
al.~\cite{ZhaiYangLiaoWrightMa2020}, who identified the global maximizers and
explicitly conjectured that all local maximizers are global---equivalently,
that the signed permutation matrices are the only local maximizers.  A related
$\ell_4$ maximization also arises in the waveform-design analysis of Liu et
al.~\cite{LiuEtAl2025OFDM}.  In the cyclic-prefix setting, their problem reduces
to the complex-unitary analogue of the square problem and only its global
maximizers are needed.  In the cyclic-prefix-free setting they study the
distinct rectangular objective
$\lVert \widetilde F_{2N}U\rVert_4^4$, prove that the DFT point is a local
maximizer, and conjecture its global optimality.  Thus that work does not prove
the absence of nonglobal local maximizers for the real square problem either.
Boumal subsequently proved the qualitative strict-saddle classification in an
online expository post~\cite{Boumal2024Blog}.

The author became aware of Boumal's blog post only after the initial version
of this manuscript had been submitted to arXiv.  The proof presented here was
developed independently.  Accordingly, we do not claim priority for the
qualitative local-maximum classification.  The presentation below emphasizes
an explicit normalized rank-two escape direction, the exact orthostochastic
pivot identity~\eqref{eq:pivot-identity}, and the quantitative curvature
bound~\eqref{eq:quantitative}, without relying on a classification of
stationary orthogonal matrices or on a relaxation to the Birkhoff polytope.

\section{Main theorem}

\begin{theorem}[Complete local-maximum classification]
For every integer $r\ge1$, let
\[
 f_r(Q)=\sum_{i,j=1}^r q_{ij}^4,
 \qquad Q\in\Or(r).
\]
Then:
\begin{enumerate}[label=\textup{(\roman*)}]
\item $f_r(Q)\le r$, with equality exactly at the signed permutation
matrices;
\item every signed permutation is a strict local maximum (with the usual
vacuous isolated-point interpretation when $r=1$);
\item if $Q$ is stationary and is not a signed permutation, then there is an
explicit nonzero skew-symmetric matrix $\Omega$ such that
\[
 \left.\frac{d^2}{dt^2}f_r\bigl(Qe^{t\Omega}\bigr)\right|_{t=0}>0.
\]
Consequently, the signed permutation matrices are exactly the local
maximizers of $f_r$ on $\Or(r)$.
\end{enumerate}
\end{theorem}

The escaping matrix $\Omega$ is constructed in
Section~\ref{sec:main-proof} from a largest entry of the squared-entry matrix
$Q^{\circ2}$.

\section{First-order condition}

The tangent space at $Q\in\Or(r)$ is
\[
 T_Q\Or(r)=\{Q\Omega:\Omega^\top=-\Omega\}.
\]
Fix a skew-symmetric $\Omega$ and use the feasible curve
\[
 Q(t)=Qe^{t\Omega}.
\]
Since $Q'(0)=Q\Omega$ and the Euclidean gradient is
$4Q^{\circ3}$,
\begin{align*}
 \left.\frac{d}{dt}f_r(Q(t))\right|_{t=0}
 &=4\ip{Q^{\circ3}}{Q\Omega}\\
 &=4\tr\!\left((Q^{\circ3})^\top Q\Omega\right)\\
 &=4\tr(A^\top\Omega),
 \qquad A:=Q^\top Q^{\circ3}.
\end{align*}
The orthogonal complement of the skew-symmetric matrices under the Frobenius
inner product is the space of symmetric matrices.  Therefore
\[
 \boxed{
 Q\text{ is stationary}
 \quad\Longleftrightarrow\quad
 A=Q^\top Q^{\circ3}\text{ is symmetric}.}
\]
This uses every tangent direction, not merely coordinate rotations.

For comparison, rotate columns $a,b$ by
\[
 q_a(t)=q_a\cos t+q_b\sin t,
 \qquad
 q_b(t)=-q_a\sin t+q_b\cos t.
\]
Direct differentiation of the affected two columns gives
\begin{align}
 f'_{ab}(0)
 &=4\sum_iq_{ia}q_{ib}(q_{ia}^2-q_{ib}^2),
 \label{eq:givens-first}\\
 f''_{ab}(0)
 &=4\sum_i
 \left(6q_{ia}^2q_{ib}^2-q_{ia}^4-q_{ib}^4\right).
 \label{eq:givens-second}
\end{align}
Equation~\eqref{eq:givens-first} vanishes for every pair at a stationary
point, consistently with the full matrix condition above.  Formula
\eqref{eq:givens-second} is only a sanity check; the proof below does not
assume that coordinate Givens directions suffice.

\section{Second-order formula}

Because $Q''(0)=Q\Omega^2$, direct differentiation gives
\begin{align*}
 f_r''(0)
 &=12\sum_{k,l}q_{kl}^2(Q\Omega)_{kl}^2
   +4\ip{Q^{\circ3}}{Q\Omega^2}.
\end{align*}
At a stationary point $A=A^\top$, and hence
\[
 \ip{Q^{\circ3}}{Q\Omega^2}
 =\tr(A\Omega^2)
 =-\tr(A\Omega^\top\Omega).
\]
Thus
\begin{equation}
 \boxed{
 f_r''(0)=4\mathcal H_Q(\Omega),
 \qquad
 \mathcal H_Q(\Omega)
 =3\sum_{k,l}q_{kl}^2(Q\Omega)_{kl}^2
  -\tr(A\Omega^\top\Omega).}
 \label{eq:full-hessian}
\end{equation}
For maximization, a second-order necessary condition at a local maximum is
$\mathcal H_Q(\Omega)\le0$ for every skew $\Omega$.  Hence any
$\Omega$ with $\mathcal H_Q(\Omega)>0$ is a strict feasible escape
certificate.

\section{Main proof: the maximum-entry pivot}
\label{sec:main-proof}

Let $Q$ be stationary.  Define
\[
 M=Q^{\circ2},\qquad
 s_i=\sum_lM_{il}^2,\qquad
 t_j=\sum_kM_{kj}^2,
 \qquad C=MM^\top M.
\]
Orthogonality of the rows and columns of $Q$ shows that $M$ is doubly
stochastic:
\[
 M_{ij}\ge0,\qquad
 \sum_jM_{ij}=1,\qquad
 \sum_iM_{ij}=1.
\]

\subsection{An exact pivot identity}

Fix indices $(i,j)$ for which
\[
 p=M_{ij}=q_{ij}^2<1,
 \qquad q=q_{ij}.
\]
Regard the transpose of row $i$ as the unit column vector
\[
 x=Q^\top e_i.
\]
Then $Qx=e_i$ and $x_j=q$.  Set
\begin{equation}
 v=\frac{e_j-qx}{\sqrt{1-p}},
 \qquad
 \Omega=xv^\top-vx^\top.
 \label{eq:pivot-direction}
\end{equation}
Indeed,
\[
 x^\top v=0,
 \qquad
 \lVert e_j-qx\rVert_2^2=1-p,
\]
so $x,v$ are orthonormal.  Thus $\Omega$ is nonzero, skew-symmetric,
rank two, and generates the feasible curve $Qe^{t\Omega}$.

Since
\[
 Qv=\frac{Qe_j-qe_i}{\sqrt{1-p}},
\]
a direct cancellation in $Q\Omega=e_iv^\top-(Qv)x^\top$ gives
\begin{equation}
 (Q\Omega)_{kl}
 =\frac{\delta_{ki}\delta_{lj}-q_{kj}q_{il}}{\sqrt{1-p}}.
 \label{eq:pivot-entry}
\end{equation}
Consequently,
\begin{align}
 \sum_{k,l}q_{kl}^2(Q\Omega)_{kl}^2
 &=\frac{C_{ij}+p-2p^2}{1-p},
 \label{eq:weighted-pivot}\\
 C_{ij}
 &=(MM^\top M)_{ij}
   =\sum_{k,l}M_{il}M_{kl}M_{kj}.
 \notag
\end{align}

It remains to compute the multiplier term in
\eqref{eq:full-hessian}.  This is the step where full stationarity is
essential.  Since $A=A^\top$,
\[
 Ax=A^\top Q^\top e_i
 =(Q^{\circ3})^\top QQ^\top e_i
 =(Q^{\circ3})^\top e_i
 =x^{\circ3}.
\]
It follows that
\[
 x^\top Ax=s_i,\qquad
 e_j^\top Ae_j=t_j,\qquad
 x^\top Ae_j=e_j^\top Ax=q^3.
\]
Moreover, $\Omega^\top\Omega=xx^\top+vv^\top$.  Therefore
\begin{equation}
 \tr(A\Omega^\top\Omega)
 =\frac{s_i+t_j-2p^2}{1-p}.
 \label{eq:multiplier-pivot}
\end{equation}
Combining \eqref{eq:full-hessian}, \eqref{eq:weighted-pivot}, and
\eqref{eq:multiplier-pivot} proves the exact identity
\begin{equation}
 \boxed{
 \mathcal H_Q(\Omega)
 =\frac{N_{ij}}{1-M_{ij}},
 \qquad
 N_{ij}=3(MM^\top M)_{ij}+3M_{ij}-4M_{ij}^2-s_i-t_j.}
 \label{eq:pivot-identity}
\end{equation}

\subsection{Strict positivity at a largest squared entry}

Assume first that every entry of $M$ is strictly below one.  Choose a largest
entry
\[
 m=M_{ij}=\max_{k,l}M_{kl}.
\]
Since $M$ is nonempty and doubly stochastic, $0<m<1$.  Write
$S=s_i+t_j$.  In
\[
 C_{ij}=\sum_{k,l}M_{il}M_{kl}M_{kj},
\]
the slice $k=i$ contributes $ms_i$, the slice $l=j$ contributes $mt_j$,
and their common term is $m^3$.  Hence the following is an exact disjoint
decomposition:
\begin{equation}
 C_{ij}=mS-m^3+R,
 \qquad
 R=\sum_{\substack{k\ne i\\l\ne j}}
 M_{il}M_{kl}M_{kj}\ge0.
 \label{eq:C-decomposition}
\end{equation}
The numerator in \eqref{eq:pivot-identity} becomes
\begin{equation}
 N_{ij}=3R+(3m-1)S+3m-4m^2-3m^3.
 \label{eq:N-base}
\end{equation}

There are two cases.

\paragraph{Case 1: $m\ge\frac13$.}
The selected row and column each contain the entry $m$, so
\[
 S=2m^2+D_+,
 \qquad
 D_+=\sum_{l\ne j}M_{il}^2+
      \sum_{k\ne i}M_{kj}^2\ge0.
\]
Substitution in \eqref{eq:N-base} gives
\begin{equation}
 N_{ij}=3R+(3m-1)D_++3m(1-m)^2>0.
 \label{eq:large-m}
\end{equation}

\paragraph{Case 2: $m\le\frac13$.}
Maximality of $m$ and double stochasticity give
\[
 s_i=\sum_lM_{il}^2\le m\sum_lM_{il}=m,
 \qquad
 t_j=\sum_kM_{kj}^2\le m\sum_kM_{kj}=m.
\]
Thus
\[
 S=2m-D_-,
 \qquad
 D_-=(m-s_i)+(m-t_j)\ge0.
\]
Now \eqref{eq:N-base} gives
\begin{equation}
 N_{ij}=3R+(1-3m)D_-+m(1-m)(1+3m)>0.
 \label{eq:small-m}
\end{equation}
At $m=1/3$, the two formulas agree.  Since $1-m>0$,
\eqref{eq:pivot-identity}--\eqref{eq:small-m} prove
\[
 \mathcal H_Q(\Omega)>0.
\]
More quantitatively, the displayed direction satisfies
\begin{equation}
 \mathcal H_Q(\Omega)\ge
 \begin{cases}
 3m(1-m),&m\ge1/3,\\
 m(1+3m),&m\le1/3.
 \end{cases}
 \label{eq:quantitative}
\end{equation}

\subsection{Entries of magnitude one}

The construction \eqref{eq:pivot-direction} is deliberately not used when
$M_{ij}=1$.  In that case $q_{ij}=\pm1$, and the unit norms of row $i$ and
column $j$ force every other entry in that row and column to vanish.  After
row and column signed permutations,
\[
 Q=(1)\oplus Q',
 \qquad Q'\in\Or(r-1).
\]
The stationarity matrix is block diagonal, so stationarity of $Q$ passes to
$Q'$.  Remove all such signed singleton blocks.  If no block remains, the
original $Q$ was a signed permutation.  Otherwise the remaining stationary
orthogonal block has a largest squared entry $m\in(0,1)$, so the construction
above gives a strict ascent direction in that block.  Extending its skew
matrix by zero gives the same strict ascent direction for the original $Q$.

We have now proved that every stationary non-signed-permutation has an
explicit $\Omega$ with $f_r''(0)=4\mathcal H_Q(\Omega)>0$.  Since the first
derivative vanishes at a stationary point, Taylor expansion gives
\[
 f_r(Qe^{t\Omega})
 =f_r(Q)+2\mathcal H_Q(\Omega)t^2+o(t^2)>f_r(Q)
\]
for all sufficiently small nonzero $t$.  Such a point cannot be a local
maximum.

\subsection{Signed permutations are the strict global maxima}

For every row of $Q$,
\[
 \sum_jq_{ij}^4
 \le\left(\sum_jq_{ij}^2\right)^2=1.
\]
Thus $f_r(Q)\le r$.  Equality holds precisely when every row has exactly one
nonzero entry, of magnitude one; column orthogonality then makes $Q$ a signed
permutation.

At $Q=I$, one has $A=I$.  Since every skew matrix has zero diagonal, the
first term in \eqref{eq:full-hessian} vanishes and
\[
 \mathcal H_I(\Omega)=-\tr(\Omega^\top\Omega)
 =-\lVert\Omega\rVert_F^2<0
\]
for every nonzero skew $\Omega$.  Left and right multiplication by signed
permutations preserves the objective and Hessian inertia, so every signed
permutation is a strict local maximum for $r\ge2$.  For $r=1$, the two points
of $\Or(1)$ are isolated and are both signed permutations.  This completes
the theorem.

\section{Edge cases and degeneracies}

The proof does not use a hidden genericity assumption.

\begin{enumerate}[label=\textup{(\alph*)}]
\item \textbf{$r=1$.}
$\Or(1)=\{[-1],[1]\}$, and both points are signed permutations with value
$1$.

\item \textbf{$r=2$.}
The general proof applies without modification.  As a direct check, an
orthogonal matrix is, up to signs and permutations, a planar rotation and
has objective $2(\cos^4\theta+\sin^4\theta)$.  Its local maxima occur at
$\theta\in(\pi/2)\mathbb Z$, exactly the signed permutations.  The
$45^\circ$ points have an ascent direction for maximization.

\item \textbf{Zero entries.}
The pivot construction never divides by $q_{ij}$.  A largest entry of the
doubly stochastic matrix $M$ is positive, and the only denominator is
$1-m$, which is positive after signed singleton blocks are removed.

\item \textbf{Repeated magnitudes.}
The largest entry need not be unique.  Any maximizing index pair satisfies
the same decomposition and strict estimate.

\item \textbf{Block diagonal and disconnected support.}
An entry of magnitude one is handled by exact block peeling.  More general
orthogonal direct sums cause no problem: select a largest squared entry in a
nontrivial residual block, build $\Omega$ there, and extend it by zero.

\item \textbf{Hadamard-type stationary points.}
If a normalized real Hadamard matrix of order $r>1$ is used, then
$M=J_r/r$, $s_i=t_j=1/r$, and $C=M$.  Formula
\eqref{eq:pivot-identity} gives
\[
 \mathcal H_Q(\Omega)=\frac4r>0.
\]
Thus these highly repeated, full-support stationary points are explicitly
excluded as local maxima.

\item \textbf{Pairwise-stable but mixed-direction-unstable points.}
The proof uses the full Hessian direction \eqref{eq:pivot-direction}, not an
assumption that elementary row or column rotations suffice.  It therefore
also handles structured points for which all coordinate Givens curvatures
are nonpositive but a mixed direction has positive curvature.

\item \textbf{Orthostochastic singularities.}
Although $M=Q^{\circ2}$ is doubly stochastic, the proof never replaces the
orthostochastic image by the Birkhoff polytope.  Double stochasticity is used
only in the elementary inequalities for $M$; the ascent direction is lifted
explicitly to the feasible curve $Qe^{t\Omega}$.
\end{enumerate}

\section{Proof scope and technical checks}

For completeness, we record how the proof addresses the standard degeneracies
that can arise in entrywise optimization on the orthogonal group.

\begin{enumerate}[label=\arabic*.]
\item \textbf{Square-only formulation:} every matrix is $r\times r$; no
rectangular Stiefel problem is introduced.
\item \textbf{Full stationarity:} symmetry of $Q^\top Q^{\circ3}$ is derived
from all skew directions, rather than inferred from an incomplete test.
\item \textbf{Second-order sign and factor:} direct differentiation confirms
$f''(0)=4\mathcal H_Q(\Omega)$; positive curvature is ascent for maximization.
\item \textbf{Feasibility:} $\Omega$ in \eqref{eq:pivot-direction} is
explicitly skew, and $Qe^{t\Omega}\in\Or(r)$ for every real $t$.
\item \textbf{Stationarity usage:} the identity $Ax=x^{\circ3}$ is invoked
only after $A=A^\top$ has been established.  It is not assumed for a generic
orthogonal matrix.
\item \textbf{Index order:} the weighted term expands as
$C_{ij}=(MM^\top M)_{ij}
=\sum_{k,l}M_{il}M_{kl}M_{kj}$, ruling out a hidden transpose error.
\item \textbf{Zeros and repeated magnitudes:} no nonzero-entry or uniqueness
assumption is made.
\item \textbf{Endpoint $m=1$:} the undefined pivot is never used there;
signed singleton blocks are removed exactly.
\item \textbf{Case split:} for $3m-1\ge0$ the lower bound
$S\ge2m^2$ is used, while for $3m-1\le0$ the upper bound $S\le2m$ is used.
The inequality direction is therefore correct in both branches.
\item \textbf{Strictness:} the residual terms $3m(1-m)^2$ and
$m(1-m)(1+3m)$ are strictly positive for $0<m<1$; the boundary
$m=1/3$ has no gap.
\item \textbf{Blocks and degeneracies:} singleton blocks, arbitrary direct
sums, normalized Hadamard points, and pairwise-insufficient examples are all
covered by the same pivot mechanism.
\item \textbf{Local versus global:} the pivot proof rules out every
non-permutation local maximum, while the separate rowwise inequality proves
global optimality.
\item \textbf{Strict-saddle wording:} the precise conclusion is
$\lambda_{\max}(\operatorname{Hess}f_r)>0$ at each non-permutation stationary
point; Hessian indefiniteness is not claimed.
\item \textbf{No circular classification or black box:} no classification of
stationary points, external database, numerical search, or theorem of
comparable difficulty is used in the proof.
\end{enumerate}


\begin{thebibliography}{9}

\bibitem{EdelmanAriasSmith1998}
A. Edelman, T. A. Arias, and S. T. Smith,
The geometry of algorithms with orthogonality constraints,
\emph{SIAM Journal on Matrix Analysis and Applications} \textbf{20} (1998),
no.~2, 303--353.
\url{https://doi.org/10.1137/S0895479895290954}

\bibitem{AbsilMahonySepulchre2008}
P.-A. Absil, R. Mahony, and R. Sepulchre,
\emph{Optimization Algorithms on Matrix Manifolds},
Princeton University Press, Princeton, NJ, 2008.

\bibitem{BanicaNechita2013}
T. Banica and I. Nechita,
Almost Hadamard matrices: the case of arbitrary exponents,
\emph{Discrete Applied Mathematics} \textbf{161} (2013), no.~16--17,
2367--2379.
\url{https://doi.org/10.1016/j.dam.2013.05.012}

\bibitem{ZhaiYangLiaoWrightMa2020}
Y. Zhai, Z. Yang, Z. Liao, J. Wright, and Y. Ma,
Complete dictionary learning via $\ell_4$-norm maximization over the
orthogonal group,
\emph{Journal of Machine Learning Research} \textbf{21} (2020), no.~165,
1--68.
\url{https://jmlr.org/papers/v21/19-755.html}

\bibitem{LiuEtAl2025OFDM}
F. Liu, Y. Zhang, Y. Xiong, S. Li, W. Yuan, F. Gao, S. Jin, and G. Caire,
CP-OFDM achieves the lowest average ranging sidelobe under QAM/PSK
constellations,
\emph{IEEE Transactions on Information Theory} \textbf{71} (2025), no.~9,
6950--6967.
\url{https://doi.org/10.1109/TIT.2025.3591267}

\bibitem{Boumal2024Blog}
N. Boumal,
Maximizing the 4-norm over orthogonal matrices has a benign landscape,
\emph{Race to the Bottom: the OPTIM@EPFL blog}, online post,
December 30, 2024; modified July 8, 2026.
\url{https://www.racetothebottom.xyz/posts/four-norm-orthogonal/}

\end{thebibliography}
\end{document}